\documentclass[twocolumn]{aastex631}%[linenumbers]{aastex631}%

\newcommand\new[1]{{\textbf{#1}}}%\color{blue}#1}}          % new text for referee

\renewcommand\new[1]{\color{black}#1}

\shorttitle{vZWFS On-Sky}
\shortauthors{Salama et al.}
\graphicspath{{./}{figures/}}
\begin{document}

\title{Keck Primary Mirror Closed-Loop Segment Control using a Vector-Zernike Wavefront Sensor}

\correspondingauthor{Ma\"issa Salama}
\email{msalama@ucsc.edu}

\author[0000-0002-5082-6332]{Ma\"issa Salama}
\affiliation{Department of Astronomy \& Astrophysics, University of California, Santa Cruz, 1156 High Street, Santa Cruz, CA 95064, USA}

\author[0009-0007-8340-6194]{Charlotte Guthery}
\affiliation{W. M. Keck Observatory, 65-1120 Mamalahoa Hwy, Kamuela, HI 96743, USA}

\author{Vincent Chambouleyron}
\affiliation{Department of Astronomy \& Astrophysics, University of California, Santa Cruz, 1156 High Street, Santa Cruz, CA 95064, USA}

\author[0000-0003-0054-2953]{Rebecca Jensen-Clem}
\affiliation{Department of Astronomy \& Astrophysics, University of California, Santa Cruz, 1156 High Street, Santa Cruz, CA 95064, USA}

\author[0000-0001-5299-6899]{J. Kent Wallace}
\affiliation{Jet Propulsion Laboratory, California Institute of Technology, 4800 Oak Grove Dr, Pasadena, CA 91109}

\author[0000-0001-8953-1008]{Jacques-Robert Delorme}
\affiliation{W. M. Keck Observatory, 65-1120 Mamalahoa Hwy, Kamuela, HI 96743, USA}

\author[0000-0002-0040-5409]{Mitchell Troy}
\affiliation{Jet Propulsion Laboratory, California Institute of Technology, 4800 Oak Grove Dr, Pasadena, CA 91109}

\author{Tobias Wenger}
\affiliation{Jet Propulsion Laboratory, California Institute of Technology, 4800 Oak Grove Dr, Pasadena, CA 91109}

\author[0000-0002-1583-2040]{Daniel Echeverri}
\affiliation{Department of Astronomy, California Institute of Technology, Pasadena, CA 91125, USA}

\author[0000-0002-1392-0768]{Luke Finnerty}
\affiliation{Department of Physics \& Astronomy, 430 Portola Plaza, University of California, Los Angeles, CA 90095, USA}

\author[0000-0001-5213-6207]{Nemanja Jovanovic}
\affiliation{Department of Astronomy, California Institute of Technology, Pasadena, CA 91125, USA}

\author[0000-0002-4934-3042]{Joshua Liberman}
\affiliation{Department of Astronomy, California Institute of Technology, Pasadena, CA 91125, USA}
\affiliation{James C. Wyant College of Optical Sciences, University of Arizona, Meinel Building 1630 E. University Blvd., Tucson, AZ 85721, USA}

\author[0000-0002-2019-4995]{Ronald A. López}
\affiliation{Department of Physics, University of California, Santa Barbara, CA 93106-9530, USA}

\author[0000-0002-8895-4735]{Dimitri Mawet}
\affiliation{Department of Astronomy, California Institute of Technology, Pasadena, CA 91125, USA}

\author[0000-0003-3165-0922]{Evan C. Morris}
\affiliation{Department of Astronomy \& Astrophysics, University of California, Santa Cruz, 1156 High Street, Santa Cruz, CA 95064, USA}

\author[0000-0002-9805-3666]{Maaike van Kooten}
\affiliation{Herzberg Astronomy and Astrophysics 5071 West Saanich Road Victoria, British Columbia V9E 2E7, Canada}

\author[0000-0003-0774-6502]{Jason J. Wang}
\affiliation{ Center for Interdisciplinary Exploration and Research in Astrophysics (CIERA) and Department of Physics and Astronomy, Northwestern University, Evanston, IL 60208, USA}

\author[0000-0002-1646-442X]{Peter Wizinowich}
\affiliation{W. M. Keck Observatory, 65-1120 Mamalahoa Hwy, Kamuela, HI 96743, USA}

\author[0000-0002-6171-9081]{Yinzi Xin}
\affiliation{Department of Astronomy, California Institute of Technology, Pasadena, CA 91125, USA}

\author[0000-0002-6618-1137]{Jerry Xuan}
\affiliation{Department of Astronomy, California Institute of Technology, Pasadena, CA 91125, USA}

\begin{abstract}
We present the first on-sky segmented primary mirror closed-loop piston control using a Zernike wavefront sensor (ZWFS) installed on the Keck II telescope. Segment co-phasing errors are a primary contributor to contrast limits on Keck and will be necessary to correct for the next generation of space missions and ground-based extremely large telescopes (ELTs), which will all have segmented primary mirrors. The goal of the ZWFS installed on Keck is to monitor and correct primary mirror co-phasing errors in parallel with science observations. The ZWFS is ideal for measuring phase discontinuities such as segment co-phasing errors and is \new{one of} the most sensitive WFS, but has limited dynamic range. 
The vector-ZWFS at Keck works on the adaptive optics (AO) corrected wavefront and consists of a metasurface focal plane mask which imposes two different phase shifts on the core of the point spread function (PSF) to two orthogonal light polarizations, producing two pupil images. This design extends the dynamic range compared with the scalar ZWFS. The primary mirror segment pistons were controlled in closed-loop using the ZWFS, improving the Strehl ratio on the NIRC2 science camera by up to 10 percentage points. We analyze the performance of the closed-loop tests, the impact on NIRC2 science data, and discuss the ZWFS measurements. 
\end{abstract}

%%%%%%%%%%%%%%%%%%%%%%%%%%%%%%%%%%%%%%%%%%%%%%%%%%%%%%%%%%%%%%%%%%%%%%

%%%%%%%%%%%%%%%%%%%%%%%%%%%%%%%%%%%%%%%%%%%%%%%%%%%%%%%%%%%%%%%%%%%%%%
%%%%%%%%%%%%%%%%%%%%%%%%%%%%%%%%%%%%%%%%%%%%%%%%%%%%%%%%%%%%%%%%%%%%%%
\section{Introduction} \label{sec:intro}

Directly imaging lower-mass and closer-in exoplanets, from mature gas giants to rocky planets in the habitable-zone of the nearest stars, will require the \new{next generation of ground-based extremely large telescopes (ELTs) and space missions}, all of which will have segmented primary mirrors due to the impracticability of making and handling a single 30-40 meter diameter piece of glass. Misalignments between the segments introduce aberrations that worsen the contrast---this effect has been identified as a primary contribution to contrast limits at Keck \citep{Ragland22}. Keck is currently the only ground-based segmented telescope with an adaptive optics (AO) system and high-contrast imaging science instruments, making it a unique tool for testing new methods for reducing these segment misalignments. This will be crucial to mitigate for the next generation of ELTs \citep{JensenClem21}.

The Zernike wavefront sensor (ZWFS) uses the phase-contrast technique, first developed for microscopy, and for which Frits Zernike was awarded the Nobel Prize in 1953. This technique consists of converting phase aberrations into intensity variations that can be measured with a detector. This is done by bringing the light \new{of a point source} to a focus onto a phase-mask consisting of a dot which offsets the phase of the light passing through it without affecting the surrounding light. By matching the diameter of the dot to the core of the point spread function (PSF), we create a reference wavefront, which interferes with the rest of the PSF pattern. At the \new{relayed} pupil plane, this interference pattern encodes the phase \new{aberrations} as intensity variations. The ZWFS is \new{one of} the most sensitive wavefront sensor, making the most optimal use of photons \citep{Guyon05,Chambouleyron21}. 

A key advantage of the ZWFS is its sensitivity to phase discontinuities, which is not the case for the commonly used Shack-Hartmann and modulated Pyramid WFSs, which measure a wavefront's gradient. Phase discontinuities can be introduced by segment co-phasing errors as well as phenomena like the low wind effect (LWE), a piston pattern that appears along the secondary mirror spiders visible under low-wind conditions \citep{NDiaye16}. The LWE was discovered by the Zernike sensor for Extremely Low-level Differential Aberrations (ZELDA; \citealt{Vigan18}) on the Very Large Telescope (VLT). There have been several successful applications for wavefront sensing in astronomy using a ZWFS. ZELDA was installed to correct non-common path aberrations (NCPAs) in the SPHERE instrument on the VLT. The ZErnike Unit for Segment phasing (ZEUS; \citealt{Dohlen06,Surdej10}) tested using a seeing-limited Zernike mask for phasing a segmented deformable mirror in the laboratory and on-sky. 

\new{The work in this paper directly follows from }\cite{vanKooten22}\new{, who} demonstrated using a ZWFS to measure segment pistons on Keck II's segmented primary mirror. \new{In this work, we demonstrate closed-loop control of the primary mirror segment pistons, using the ZWFS, which resulted in improved image quality on the science camera. In addition, the Zernike mask used in \cite{vanKooten22} was replaced by a vector-Zernike mask (see Section \ref{sec:vZWFS}) to increase the dynamic range.}

The ZWFS is also being tested and considered for space-based applications. The baseline architecture for Habitable Worlds Observatory (HWO) is a 6-m segmented telescope, which will require maintaining the alignment of these segments to picometer levels to reach the sensitivity of $10^{-10}$ contrast needed to directly image an Earth-like planet around a Sun-like star \citep{LUVOIR19}. The ZWFS was baselined in the HabEx and LUVOIR mission concepts, and is being considered for the HWO \citep{Ruane23}. The ZWFS has been demonstrated to reach sensitivities down to the picometer level in experiments \citep{Steeves20,Ruane20} performed in temperature-stabilized laboratory environments. \new{Two Zernike WFS masks are} also installed on the Nancy Grace Roman Space Telescope\new{, one} to serve as a low-order wavefront sensor on the rejected starlight by the coronagraph\new{, and another as a fully transmissive ZWFS} \citep{Shi16,Riggs21}.

%%%%%%%%%%%%%%%%%%%%%%%%%%%%%%%%%%%%%%%%%%%%%%%%%%%%%%%%%%%%%%%%%%%%%%
%%%%%%%%%%%%%%%%%%%%%%%%%%%%%%%%%%%%%%%%%%%%%%%%%%%%%%%%%%%%%%%%%%%%%%

\section{\new{Vector-Zernike Wavefront Sensor}}
\label{sec:vZWFS}

\begin{figure*}
\begin{center}
\begin{tabular}{c}
\includegraphics[width=0.8\textwidth]{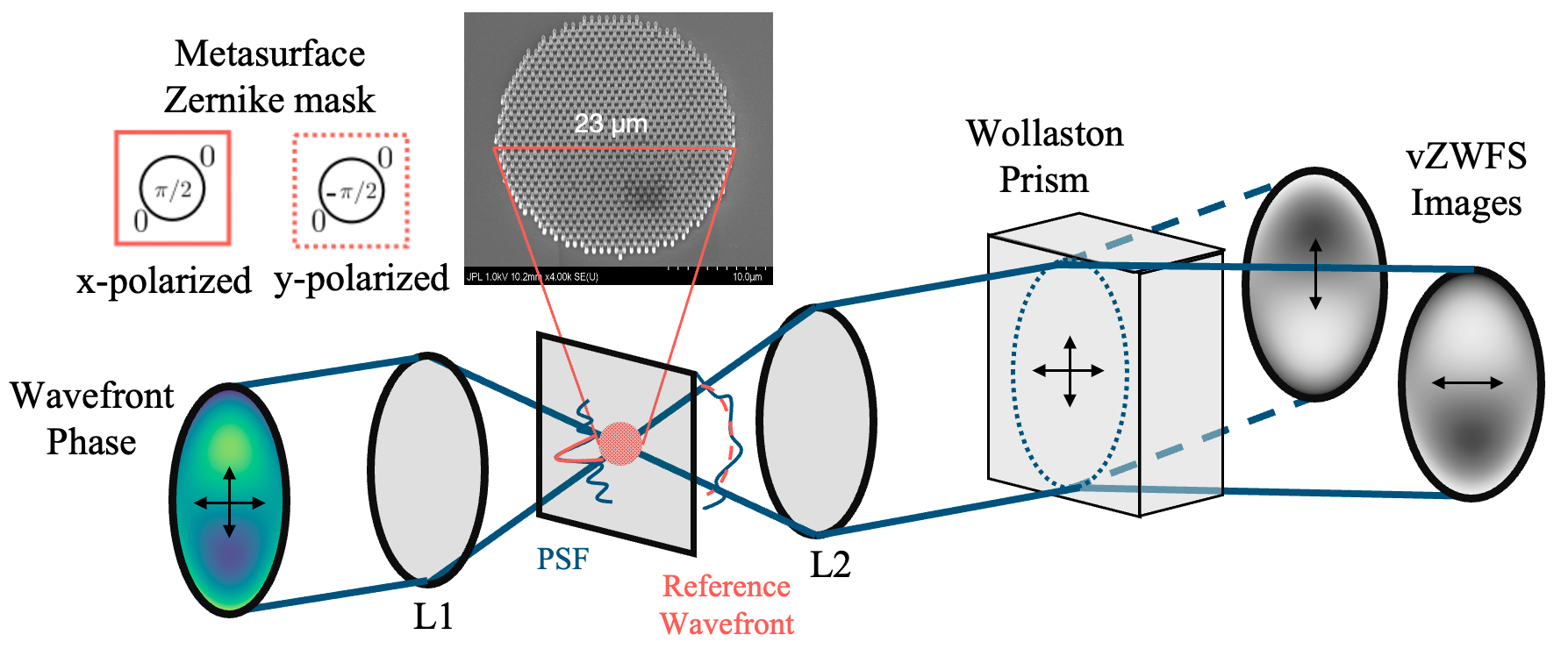}
\end{tabular}
\end{center}
\caption 
{ \label{fig:vZWFS}
Figure inspired from \cite{Doelman19} of the vector-Zernike WFS concept, applying two different phase shifts on two orthogonal light polarizations. A Wollaston prism then splits the polarizations into two images. At the top is a scanning-electron microscope image of the metasurface Zernike mask used in this study (see \S\ref{sec:metasurface}).}
\end{figure*}

Although the classical ZWFS is extremely sensitive, its dynamic range is very small. \new{The monotonic response regime is $\lambda/2$, equivalent to $\pi$ in phase, and the signal inverts beyond that \citep{NDiaye13}}. To address this central limitation of the classical ZWFS, we use a vector-Zernike wavefront sensor (vZWFS), which consists of imposing two different phase offsets simultaneously to two orthogonal polarizations, allowing us to \new{double} its dynamic range \new{to the full $2\pi$ range} through phase diversity \new{\citep{Doelman19,Cisse22}}. Figure \ref{fig:vZWFS} shows the optical layout: the light is focused by a converging lens (L1) onto the Zernike mask, which applies a $+\pi/2$ phase offset to the x-polarized PSF core and a $-\pi/2$ phase offset to the y-polarized PSF core. A second lens (L2) then collimates the beam, which then goes through a Wollaston prism, splitting the x and y polarized light into two beams, resulting in two pupil images. 

%%%%%%%%%%%%%%%%%%%%%%%%%%%%%%%%%%%%%%%%%%%%%%%%%%%%%%%%%%%%%%%%%%%%%%
\subsection{Phase Reconstruction}
\label{sec:PhaseRecon}

We reconstruct the phase following the method detailed in \cite{Chambouleyron23} and based on \cite{Steeves20}, using a numerical model of the system and iterative algorithm for a non-linear reconstruction. To reconstruct the phase for a ZWFS with a dimple phase-shift of $\theta_{0}$ radians, we use the following formula:
\begin{equation}
\phi(\textbf{x}) = \beta(\textbf{x}) + \frac{\theta_{0}}{2} + \arcsin \bigg[ \frac{I(\textbf{x})-4\cdot|\sin (\frac{\theta_{0}}{2})|^{2}\cdot \textit{I}_{b(\textbf{x})}-\textit{I}_{P(\textbf{x})}}{4\cdot \sin (\frac{\theta_{0}}{2})\cdot \sqrt{\textit{I}_{b(\textbf{x})} \textit{I}_{P(\textbf{x})}}}\bigg]
\label{eq:phase}
\end{equation}
where $\phi$ is the wavefront phase estimation \new{at each pixel location \textbf{x}}, $\textit{I}$ is the ZWFS image and $\textit{I}_{P}$ is the pupil image. $\textit{I}_{b}$ and $\beta$ are respectively the image and the phase of the reference wave, which corresponds to the electric field going through the ZWFS dimple only. These quantities are computed and updated \new{iteratively in the reconstruction algorithm. For each iterative step, the phase estimation is propagated through the numerical model of the ZWFS to produce a new version of the reference wave, allowing the change in Strehl ratio seen by the mask to be taken into account.} The two pupil images are individually reconstructed according to their phase-shifts, then their reconstructed phases are combined through pixel by pixel matrix inversion in order to exploit the phase diversity brought by the vZWFS configuration. More details, as well as an evaluation of this and other ZWFS phase reconstruction methods, are the subject of a forthcoming article \citep{Chambouleyron24}.

%%%%%%%%%%%%%%%%%%%%%%%%%%%%%%%%%%%%%%%%%%%%%%%%%%%%%%%%%%%%%%%%%%%%%%
%%%%%%%%%%%%%%%%%%%%%%%%%%%%%%%%%%%%%%%%%%%%%%%%%%%%%%%%%%%%%%%%%%%%%%

\section{Setup at Keck}
\label{sec:keck}

\begin{figure*}
\begin{center}
\begin{tabular}{c}
\includegraphics[width=0.8\textwidth]{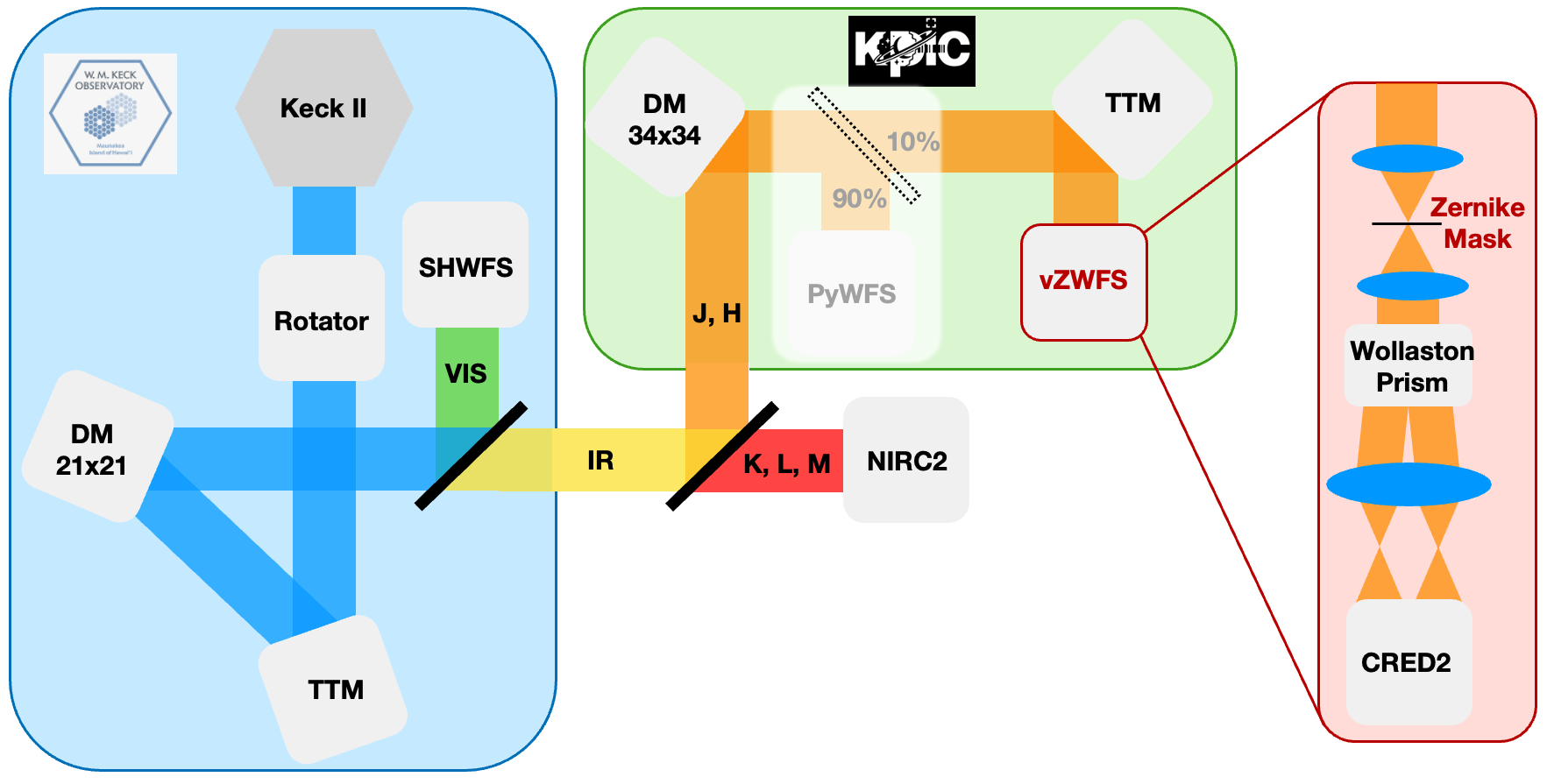}
\end{tabular}
\end{center}
\caption 
{ \label{fig:KeckSetup}
General schematic of the light path from the Keck II AO bench to the \new{vector-}Zernike wavefront sensor in the KPIC instrument (simplified here to only the vZWFS-relevant components). The PyWFS is shown for reference, but was not used for the results presented in this work. The beam splitter that normally sends 90\% of the light to the PyWFS was not in the beam and thus 100\% of the light was sent to the vZWFS.}
\end{figure*}

The vZWFS was installed as part of the Keck Planet Imager and Characterizer (KPIC; \citealt{Mawet16,Delorme21}) Phase II upgrade performed in June 2022 (\citealt{Jovanovic20,Echeverri20}, and in preparation article \citealt{Jovanovic24}). The installation of the ZWFS on Keck is part of several high-contrast tools being developed on Keck II with the goal of improving PSF quality and achievable contrasts in parallel with science observations \citep{Guthery23}. The vector-Zernike mask replaced the scalar Zernike mask used in \cite{vanKooten22}. Figure \ref{fig:KeckSetup} shows a schematic of the light path from the telescope to the vZWFS. The light enters the Keck II telescope, the rotator sends it to the AO system \citep{Wizinowich00}, which includes a tip-tilt mirror (TTM) and a 21$\times$21 actuator deformable mirror (DM). A dichroic sends the visible light to the Shack-Hartmann WFS (SHWFS), the K, L, M, near-infrared wavelengths are sent to the NIRC2 science camera, and the J and H near-infrared wavelengths are sent to the KPIC instrument. Within KPIC, there is a 34x34 DM, which is used by the vZWFS to correct for NCPAs. If the Pyramid WFS (PyWFS; \citealt{Bond20}) is used instead of the SHWFS for the AO correction, then 90\% of the J and H-band light is sent to it and the remaining 10\% of the light is let through. For this study we operated using the SHWFS to close the AO loop\footnote{The dichroic that sends light to the PyWFS causes a polarization effect, essentially removing one of the polarizations and thus we lose one of our pupil images. This will be mitigated by adding a half-wave plate before the vector-Zernike mask during the upcoming KPIC Service Mission in early 2024.} and therefore removed the dichroic from the beam so that 100\% of the light continues through to a TTM which centers the PSF onto the Zernike mask. The light is then collimated, the two orthogonal polarizations are split into two beams by the Wollaston prism, \new{a final lens reimages the pupils,} and finally the resulting two pupil images are imaged on the CRED2 detector with 202-pixel diameters each. \new{Also see \cite{Wallace23} for a detailed description of the vZWFS optical layout at Keck.}

\begin{figure}
    \centering
    \includegraphics[width=0.45\textwidth]{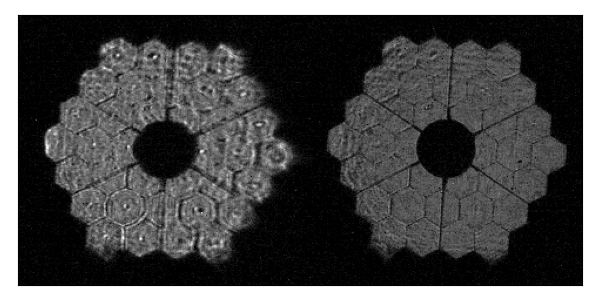}
    \caption{Reference pupil intensity images taken without going through the Zernike mask. The difference in focus in the two pupil images is due to the different path lengths from the Wollaston prism splitting the polarized light.}
    \label{fig:OffMask}
\end{figure}

A drawback of this mask in this setup is the focus difference between the two pupil images due to the path difference introduced by the Wollaston prism. \new{As seen in Figure \ref{fig:OffMask}, the left pupil image is out of focus. This effect impacts the details of the high-spatial scale structures in the reconstructed wavefront, as the two pupil images are combined when reconstructing the phase (Section \ref{sec:PhaseRecon}). However, the} phase diversity brought by the two phase shifts still allows for an increased dynamic range \new{ and for reconstructing segment pistons, which are mid-spatial scale structures.}

%%%%%%%%%%%%%%%%%%%%%%%%%%%%%%%%%%%%%%%%%%%%%%%%%%%%%%%%%%%%%%%%%%%%%%
\subsection{Metasurface vector-Zernike mask} 
\label{sec:metasurface}

To extend the dynamic range of the scalar phase-contrast technique, we implemented a vector Zernike mask that imposes two different phase shifts for the two orthogonal, linear polarization states at the Zernike dimple. The vector mask is made using metasurface optics \new{\citep{Wallace23}}. Metasurfaces are 2D arrays of subwavelength features (called nanopost or nanopillars). For a given choice of dielectric material, the nanopost geometry is optimized to produce the requisite phase shift for the two polarizations. These metasurface masks were designed at JPL, fabricated at JPL's Microdevices Laboratory, and consist of amorphous silicon nanopillars fabricated on a fused silica substrate. The Zernike dot-diameter is 23.5$\mu$m, corresponding to 2.4$\lambda$/D at $\lambda = 1600$~nm, and inscribed Keck telescope diameter, $D = 9.96$~m. The vector-Zernike mask installed on Keck and used in this study was found to have different phase-shifts than theoretically predicted. This can be due to a difference in the index of refraction of amorphous silicon used in simulations and that obtained in the fabrication process. The phase-shifts of the mask used in this study were estimated by poking an actuator on the KPIC DM over a range of amplitudes and matching the vZWFS signal response in each pupil with the simulated responses corresponding to different phase-shifts. Therefore, the phase-shifts of the mask used in this study are estimated to be 0.3$\pi$ and 0.68$\pi$ \new{(instead of the desired $\pm 0.5 \pi$). This mask was issued from the first batch of fabrication, and the fabrication process is actively being refined in order to reach the desired phase shifts. A second fabrication batch yielded masks with the desired $\pi$ difference between the two phase shifts, $0.3\pi$ and $-0.7\pi$, but not yet centered around zero.} %\todo{address referee's comment} 

\begin{figure*}
    \centering
    \includegraphics[width=0.8\textwidth]{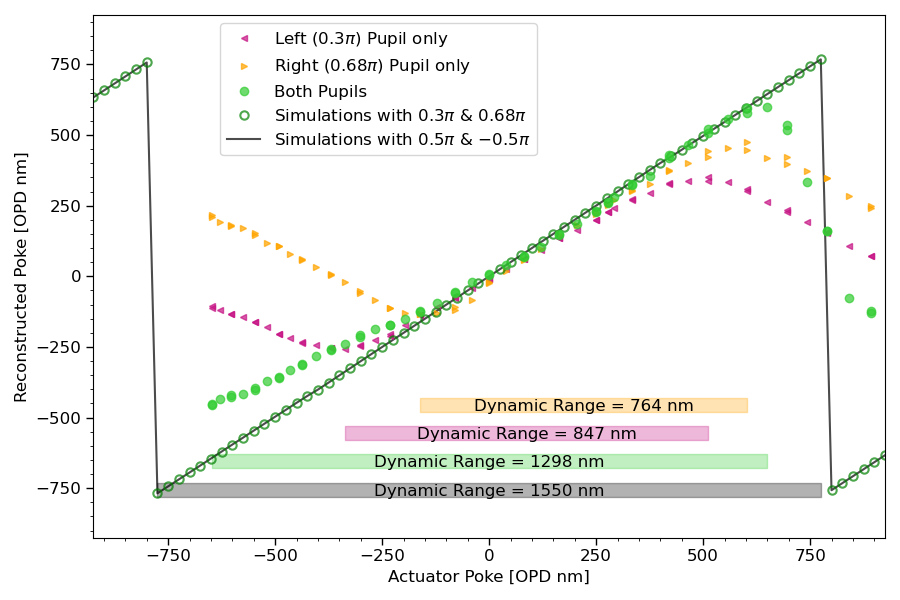}
    \caption{\new{Dynamic range measured by poking a single actuator on the KPIC DM and using the internal light source. The pink left triangles show the reconstructed poke from using only the signal in the left pupil (phase shift = 0.3$\pi$) and the pink bar is the corresponding dynamic range, the region where the signal is monotonic before it inverts. The orange right triangles correspond to only using the right pupil (phase shift = 0.68$\pi$) and green corresponds to using both pupils for the phase reconstruction. The green dynamic range is much larger than each individual pupil. The black line shows the simulated vZWFS response curve for two pupils with $\pm 0.5 \pi$ phase shifts and the green open circles show the simulated vZWFS response curve for two pupils with 0.3$\pi$ and 0.68$\pi$ phase shifts. The simulated response curve is identical. Note that the measured data does not extend beyond $-650$~nm of actuator poke, which is before the signal inversion of the two-pupil reconstruction (solid green dots) and we can thus expect the dynamic range to be slightly larger than shown.}}
    \label{fig:dynamic_range}
\end{figure*}

\new{We measured the dynamic range of this mask using the internal KPIC light source, poking the central actuator on the KPIC DM over a range of amplitudes, and reconstructing the poke amplitude from the vZWFS images, taken in the $\lambda_c = 1550$~nm narrowband ($\Delta\lambda = 25$~nm) filter. Figure \ref{fig:dynamic_range} shows the vZWFS gain in dynamic range from using both pupils, compared to the scalar ZWFS, corresponding to the phase reconstructed with just one pupil. In theory, the dynamic range of a ZWFS from a single phase shift and pupil is $\lambda/2$, and with two phase shifts and two pupils it is extended to $\lambda$. At $\lambda_c = 1550$~nm, this corresponds to 875~nm and 1550~nm, respectively. Our measurements with the internal light source are in close agreement with these theoretical predictions. With the current mask and our measurements, the dynamic range nearly doubles when using both pupils for the phase reconstruction. In addition, we illustrate what the simulated dynamic range would be if the two pupils were the designed $\pm 0.5 \pi$ phase shifts.} Since we retain the phase diversity from the two different phase shifts, we also retain our gain in dynamic range. 

%%%%%%%%%%%%%%%%%%%%%%%%%%%%%%%%%%%%%%%%%%%%%%%%%%%%%%%%%%%%%%%%%%%%%%
\subsection{Non-Common Path Calibrations}

The standard Keck AO calibrations correct NCPAs between the SHWFS and the science path by performing image sharpening of the PSF on the NIRC2 science camera. However, because the vZWFS and NIRC2 paths have different NCPAs, this PSF sharpening process introduces aberrations in the vZWFS arm. It is critical to correct for NCPAs along the vZWFS path due to its limited dynamic range. We therefore use the vZWFS to correct for NCPAs between the vZWFS path and NIRC2 by closing the loop with the vZWFS on the 34x34 DM in KPIC, which is not seen by NIRC2 \new{(see Figure \ref{fig:KeckSetup})}. The resulting shape on the DM is used as a static map on-sky. This calibration procedure is done during the day using the Keck AO calibration source. \new{The NCPA correction map generated by the vZWFS is different each time the Keck AO calibrations are re-run (which is generally each day that AO is on-sky) and the NCPAs are typically on the order of 100~nm RMS.}

%%%%%%%%%%%%%%%%%%%%%%%%%%%%%%%%%%%%%%%%%%%%%%%%%%%%%%%%%%%%%%%%%%%%%%
%%%%%%%%%%%%%%%%%%%%%%%%%%%%%%%%%%%%%%%%%%%%%%%%%%%%%%%%%%%%%%%%%%%%%%
\section{Measuring Primary Mirror Segment Co-Phasing Errors}

\subsection{On-Sky vZWFS Measurement Procedure}

We began our on-sky procedure by closing the AO loop with the SHWFS. All data was collected with the pupil at a fixed rotation angle, so that the pupil image would not rotate with telescope tracking. We first took a reference pupil image by moving the tip-tilt mirror in KPIC to offset the PSF from the Zernike mask. This provided us with reference intensities for the pupil image ($I_{P(\textbf{x})}$ term in Equation \ref{eq:phase}). We then took our first set of vZWFS measurements with the PSF aligned on the Zernike mask. We averaged out the residual atmospheric turbulence over 30-seconds, then reconstructed the phase to recover static aberrations.

To extract the individual segment piston, tip, and tilt (PTT) values from the reconstructed phase, we first removed the first few global Zernike modes (piston, tip, tilt) from the reconstructed phase. This is to avoid confusing global modes for individual segment PTT misalignments and global PTT are not modes we want to correct with the primary mirror. %We expect the wavefront reconstruction to recover some varying global tip/tilt due to the alignment of the PSF onto the mask changing as the pupil is fixed but the field rotates.

\subsection{Reconstructing Primary Mirror Segment Pistons}
\label{sec:ReconPistons}

\begin{figure*}
    \centering
    \includegraphics[width=0.7\textwidth]{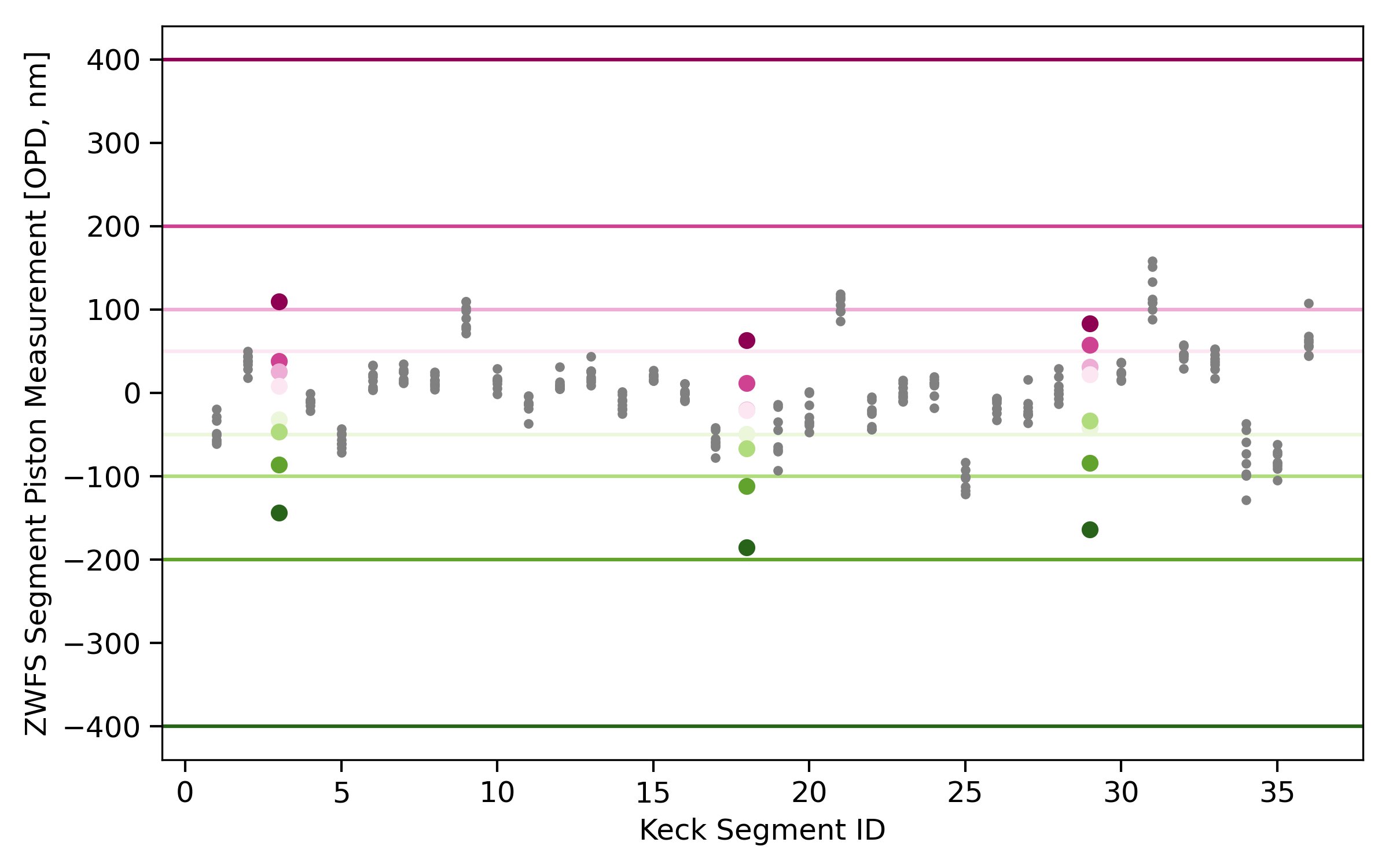}
    \caption{Reconstructed segment piston values after poking three segments over a range of piston amplitudes. Segments 3, 18, and 29 were pistoned to the levels of the horizontal lines, and the same colored points were the corresponding reconstructed pistons from the vZWFS measurements. We underestimate the phase.}
    \label{fig:PokedSegs}
\end{figure*}

To confirm our sensitivity to segment pistons, we poked segments on the Keck primary mirror by known amounts of piston. We poked segments ID 3, 17, and 29 by $\pm$50, $\pm$100, $\pm$200, and $\pm$400 nm in wavefront optical path difference (OPD). \new{The three segments were chosen as non-adjacent segments, each in a different segment ring, and were poked simultaneously by the same piston amplitude.} The segment piston measurements recovered are shown in Figure \ref{fig:PokedSegs}. The three poked segments clearly stand out from the non-poked segments. However, the recovered piston value is underestimated compared to the amplitude of the poke, by factors ranging from $\times2$ to $\times4$. This can be explained by different factors: (i) Although we are using a non-linear reconstructor, \new{our reconstructor starts to underestimate the phase below SR~$=50\%$ (at the sensing wavelength). Residuals from the AO and other static or quasi-static aberrations are therefore} taking up some of our dynamic range. (ii) \new{The reconstructor works best on short exposure frames, for which it is possible to assume coherence in order to propagate the phase estimation in the vZWFS model (Equation \ref{eq:phase}). For flux and computational time reasons, we process the vZWFS images by using average frames on the camera, which then impact reconstruction accuracy.} (iii) Potential model errors in the numerical model of the vZWFS could also impact the phase estimation. However, since we are operating in slow closed-loop iterations, these effects leading to underestimating the phase are not a major problem for wavefront correction but remain relevant for interpreting the primary mirror piston values measured by the vZWFS.

%%%%%%%%%%%%%%%%%%%%%%%%%%%%%%%%%%%%%%%%%%%%%%%%%%%%%%%%%%%%%%%%%%%%%%
%%%%%%%%%%%%%%%%%%%%%%%%%%%%%%%%%%%%%%%%%%%%%%%%%%%%%%%%%%%%%%%%%%%%%%
\section{Correcting Primary Mirror Co-Phasing Errors in Closed-Loop}
\label{sec:ClosedLoop}

\subsection{On-Sky Closed-Loop vZWFS Procedure}

Individual segment piston, tip, and tilt can be adjusted by three actuators through the primary mirror Active Control System (ACS; \citealt{Cohen94}). The ACS consists of capacitive displacement sensors which measure the relative height difference between two adjacent segment edges and sends commands to the actuators to maintain the segments within predetermined sensor alignment readings. After extracting the segment piston values from our vZWFS measurements, as explained in the previous section, we applied our desired corrections by sending segment actuator offset values to the ACS. ACS updates the desired sensor values, based on the actuator offsets, in order to implement the desired piston changes in closed loop. We then took a new set of vZWFS measurements. This was repeated for several \new{(typically 5-6)} iterations until we no longer measured an improvement in the root-mean-square (RMS) of the segment piston values. The resulting final primary mirror segment edge sensor readings were then saved in order to later easily alternate between the initial primary mirror shape and the shape generated by our vZWFS closed-loop test. We also took NIRC2 images in parallel with the vZWFS measurements so that we could directly measure and quantify the impact of the vZWFS-primary mirror closed loop on the science data, summarized in Section \S\ref{sec:science}.

\subsection{On-Sky vZWFS Closed-Loop Results}

We used the vZWFS measurements to control the primary mirror segment pistons in closed-loop on three separate nights (August 2nd, 4th, and 13th, 2023, UT) for a total of four closed-loop tests, summarized in Table \ref{tab:obs}. We closed the loop with the vZWFS on the primary mirror for the first time on August 2$^{nd}$, 2023 (UT) on two stars at two different elevations. Three segments had been exchanged during the day and had been phased (aligned in piston, tip and tilt) during the first part of the night. However, the three exchanged segments had not yet been warped to the desired surface shape. The warping of the segment shape can only be done during the day after the segment shape is measured at night. The three recently exchanged and not-yet-warped segments were IDs 13, 28, and 29. Since the segments had been phased and we were only controlling segment piston, this did not affect our closed-loop test.

\begin{table*}
    \centering
    \begin{tabular}{c|c|c|c|c|c }
        Date (UT) & Star & H-mag & Elevation range [deg] & Pupil Rotation [deg] & Seeing range \\
        \hline
        2023-08-02 & SAO68615 & 4.72 & $40 - 44$ & 0 & $0.52-0.86\arcsec$ \\
        2023-08-02 & HD221914 & 6.03 & $76 - 84$ & 0 & $0.44-1.33\arcsec$ \\
        2023-08-04 & SAO69743 & 6.94 & $77 - 78$ & 0 & $0.38-0.59\arcsec$ \\
        2023-08-13 & HIP83083 & 5.47 & $82 - 85$ & 3.73 & $0.66-0.97\arcsec$ \\
    \end{tabular}
    \caption{Summary of observing conditions during four closed loop tests. The seeing measurements are reported by the CFHT DIMM seeing monitor. See Figures \ref{fig:ClosedLoop_SRs_02} - \ref{fig:ClosedLoop_SRs_13} for a more detailed visualization of the seeing variations throughout the closed-loop runs.}
    \label{tab:obs}
\end{table*}

\begin{figure*}
    \centering
    \includegraphics[width=0.95\textwidth]{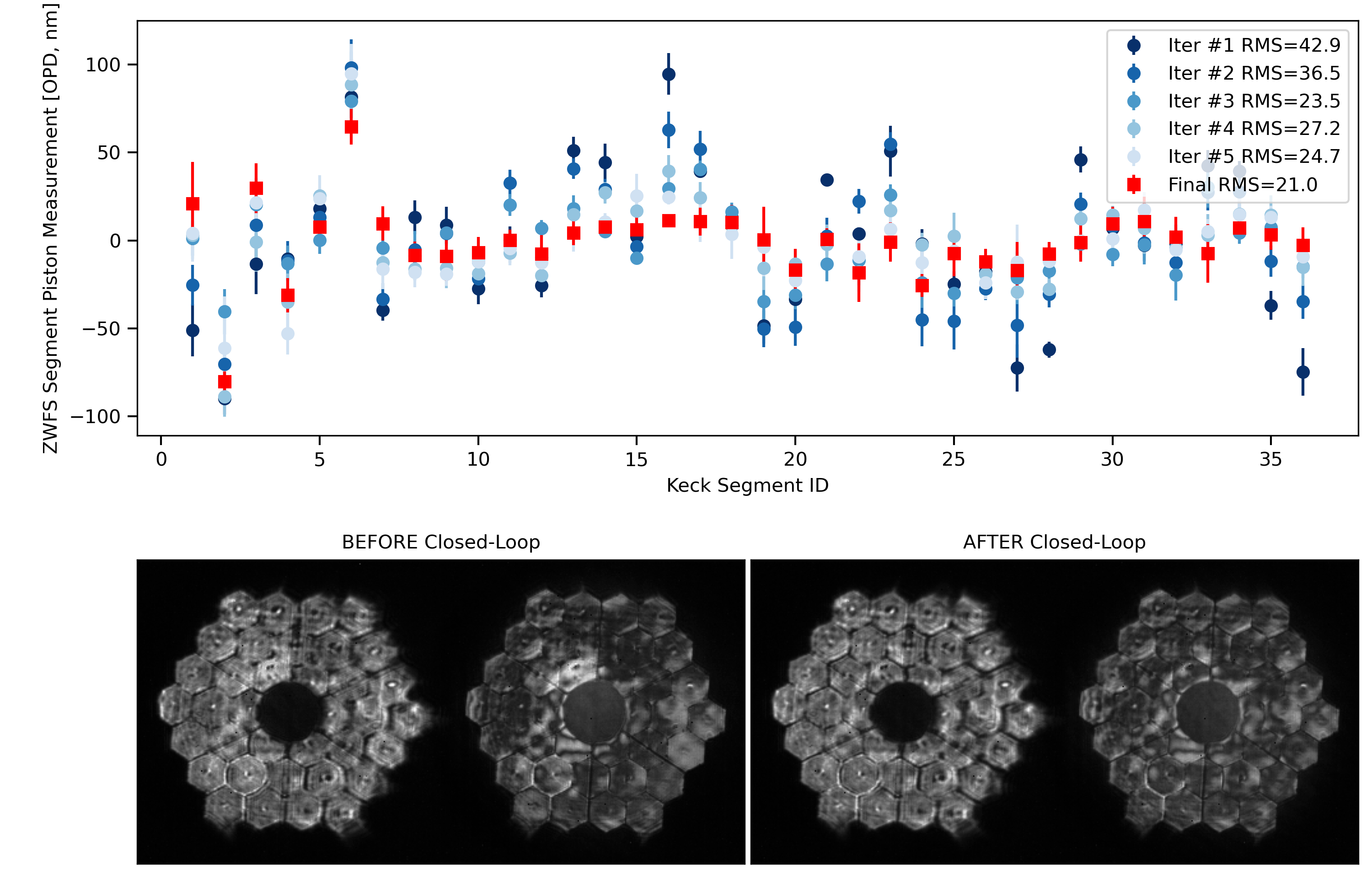}
    \caption{\textit{Top}: Mean and standard deviation piston measurement for each segment for each dataset (5 measurements) during the closed-loop iterations. The darkest blue dataset is the first iteration with subsequent iterations being lighter blues. The final post-closed loop measurement are the red squares. The RMS value of the segment pistons is decreasing with each iteration. All values are in nm of OPD. \textit{Bottom}: vZWFS images\new{, displayed in linear stretch,} before (left) and after (right) the closed-loop test. For a discussion of the higher-order spatial scale features seen in these images, see \S\ref{sec:HighSpatialScale}.}
    \label{fig:230802_CL}
\end{figure*}

The first closed-loop test was run on SAO68615 ($m_H = 4.721$) at low telescope elevation, decreasing from 44 to 40 degrees as the star was setting. The seeing ranged from $0.52-0.86\arcsec$ throughout the closed loop iterations. Figure \ref{fig:230802_CL} shows the before and after closed loop vZWFS images and the measured segment piston values for each iteration of the closed loop test. We can clearly see a reduction of the relative RMS values as the iterations progressed, though the absolute RMS values are underestimated (see \S\ref{sec:ReconPistons}). 

On the same night of August 2$^{nd}$, 2023 (UT), we ran a second closed-loop test on a fainter star at high elevation, HD221914 ($m_H = 6.03$). The seeing was on average ${\sim}0.5 - 0.75 \arcsec$, but spiked during our closed loop operations to 1.33$\arcsec$. During this spike, the AO correction was less steady, as seen in Figure \ref{fig:ClosedLoop_SRs_02}, this corresponded to a large variability in SRs on the NIRC2 PSF images taken simultaneously. This variability caused the signal on the vZWFS images to also vary by large amounts, we therefore discarded this iteration measurement and did not send the correction commands to the primary mirror. The telescope elevation changed from 84 to 76 degrees as the star was near transit. These two closed loop instances allow us to compare the performance as a function of telescope elevation (see \S\ref{sec:Elevation} and Figure \ref{fig:HighLowDiffs}). 

The next closed-loop test was on August 4$^{th}$, 2023 (UT) on SAO69743 ($m_H = 6.938$) and the telescope elevation was $\sim$77 degrees throughout the closed-loop run. The seeing was very stable during our observations, staying around $0.4-0.6\arcsec$. The telescope had been fully phased and the recently exchanged segments had been warped prior to this closed-loop test. 

On August 13$^{th}$, 2023 (UT), we closed the loop on HIP83083 ($m_H = 5.469$). The star was again at high-elevation (82 - 85 degrees) and the seeing ranged from $0.66 - 0.97 \arcsec$. However, on this night we operated prior to science observations that used the vortex coronagraph on NIRC2, which requires the pupil rotation angle to be set to 3.73 degrees. We therefore ran the vZWFS in closed-loop at this same angle, whereas all previous vZWFS closed-loop tests were carried out with a pupil rotation angle of zero \new{(see Section \S\ref{sec:OtherEffects} for a discussion on the potential impact of pupil rotation)}.

\begin{figure*}
    \centering
    \includegraphics[width=\textwidth]{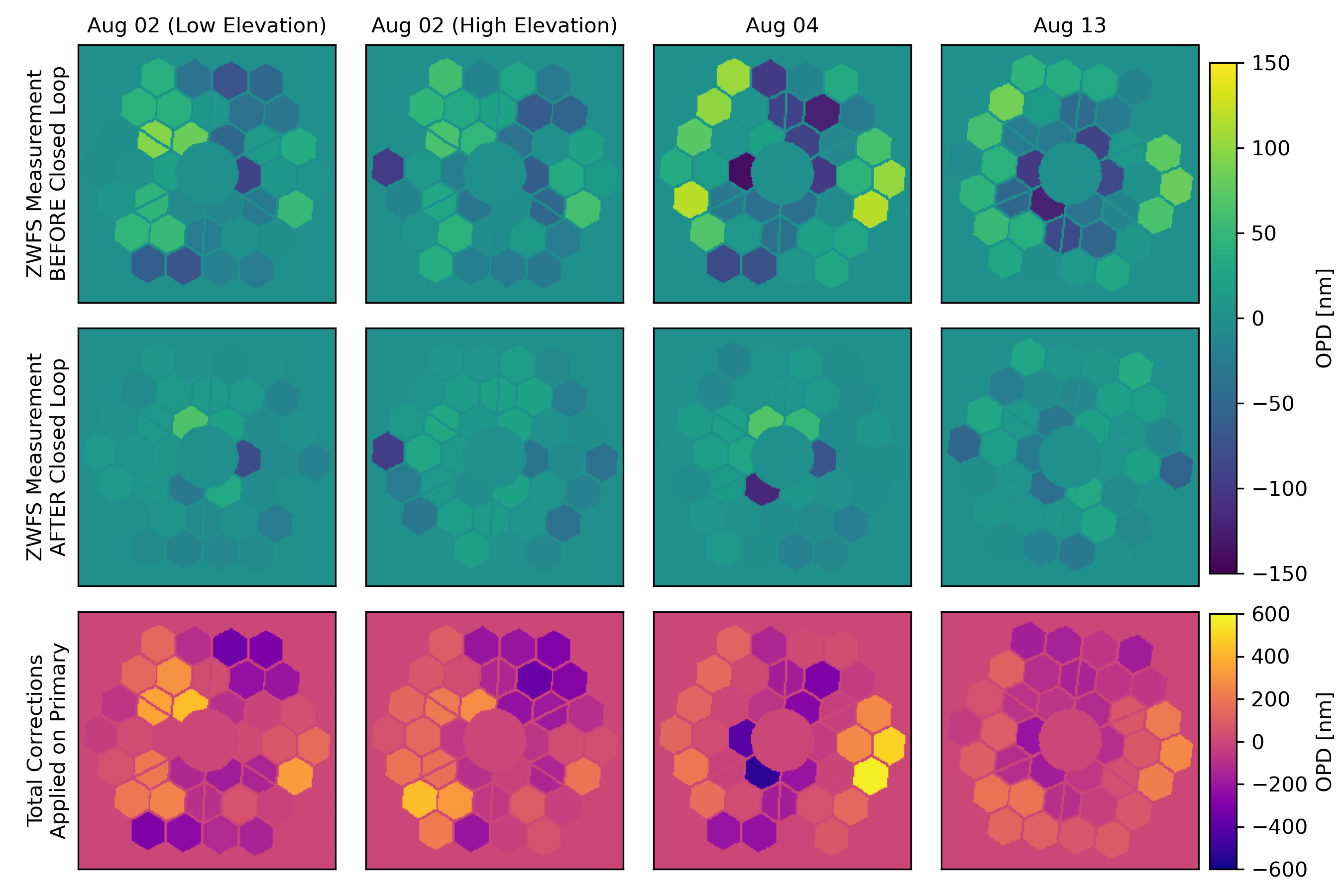}
    \caption{\textit{Top:} Segment piston measurements by the vZWFS before running the vZWFS closed-loop test. \textit{Middle:} Segment piston measurements by the vZWFS after running the vZWFS closed-loop. \textit{Bottom:} Total piston commands, in OPD nm, sent to the primary mirror segments by the end of each closed-loop run.}
    \label{fig:CL_maps}
\end{figure*}

Figure \ref{fig:CL_maps} shows the segment piston values measured by the vZWFS before and after the closed-loop runs, as well as the total piston corrections sent to the primary mirror segments by the end of the closed-loop test. Since our vZWFS measurements underestimate the phase, the total corrections sent to the primary mirror throughout the closed loop test provide us with the magnitude of the difference in the primary mirror shape.

%%%%%%%%%%%%%%%%%%%%%%%%%%%%%%%%%%%%%%%%%%%%%%%%%%%%%%%%%%%%%%%%%%%%%%
\subsection{Impact on Science Data}
\label{sec:science}

We monitored the impact of the vZWFS closed-loop tests on the NIRC2 science data by calculating the Strehl ratio (SR) in the PSF images. The SR throughout this paper was measured following Method 7 in \cite{Roberts04}. A perfect diffraction-limited PSF is generated from the Keck pupil and the ratio of the peak normalized by the total flux in a circle of radius 0.4$\arcsec$ between the real and perfect PSFs yields the SR.

\begin{figure*}
    \centering
    \includegraphics[width=\textwidth]{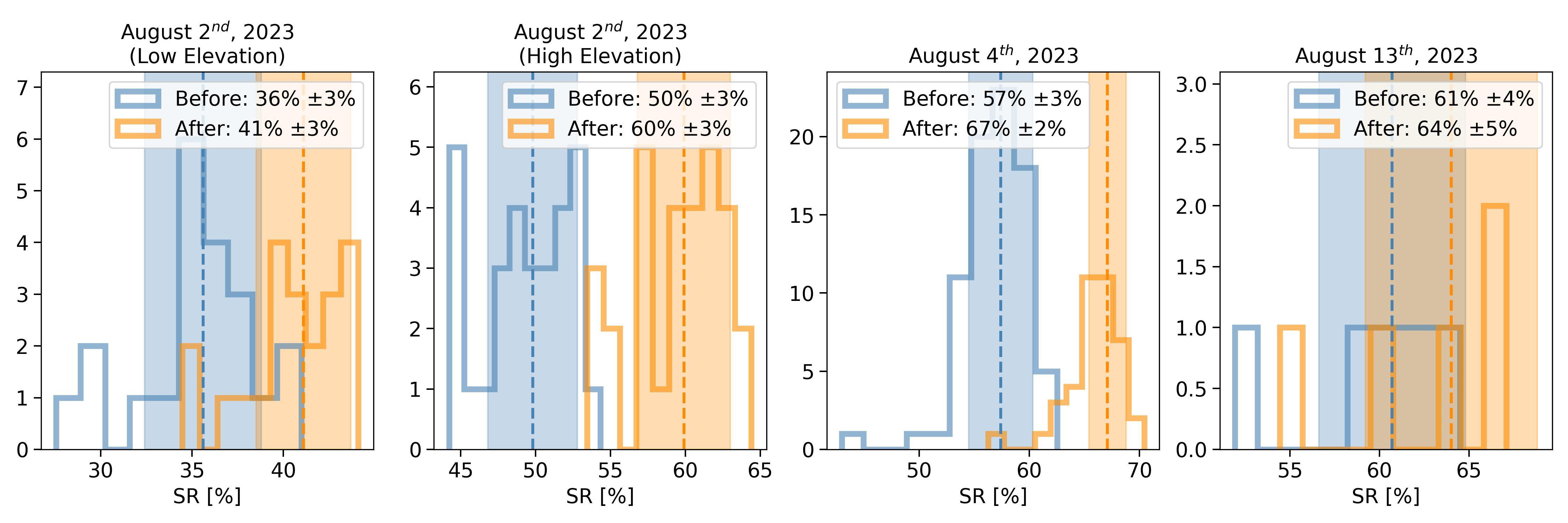}
    \caption{Histogram of NIRC2 Strehl ratios from the four closed-loop tests (blue is before and orange is after) reported in Table \ref{tab:delta_SRs}. The dashed vertical lines indicate the mean and the shaded regions represent $\pm 1 \sigma$.}
    \label{fig:CL_SRs}
\end{figure*}

\begin{figure*}
    \centering
    \includegraphics[width=\textwidth]{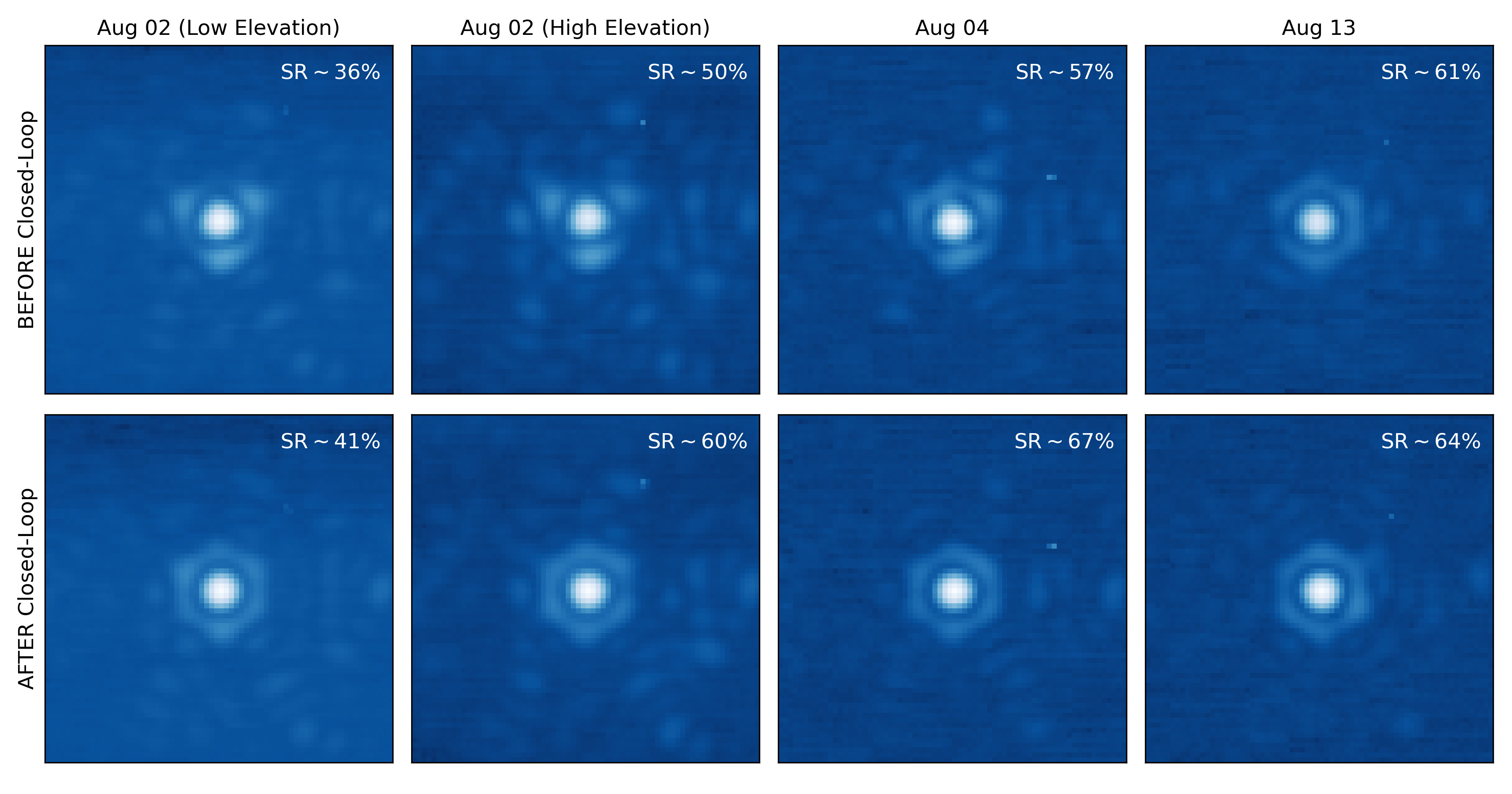}
    \caption{Mean of NIRC2 PSF images \new{in the Brackett-Gamma ($\lambda_c = 2.168 \mu m$) filter} taken before and after the four vZWFS closed-loop runs controlling the primary mirror segments. The images are displayed in square-root stretch and the color bar range match between each pair of before and after closed-loop run. The Strehl ratios are the mean SRs reported in Table \ref{tab:delta_SRs}.}
    \label{fig:CL_PSFs}
\end{figure*}

We took NIRC2 PSF images in the Brackett-Gamma ($\lambda_c = 2.168 \mu m$) filter simultaneously throughout the closed-loop iterations. After performing the closed-loop test, we alternated between the original nominal primary mirror shape and our new vZWFS-generated primary mirror shape to compare the SRs measured on the NIRC2 PSFs to ensure that changes in our measurements were not due to changing environmental conditions during the night. Table~\ref{tab:delta_SRs} \new{and Figure \ref{fig:CL_SRs} show} the SRs corresponding to images taken with the nominal and the vZWFS-generated primary mirror shapes. This shows 3 out of 4 closed loop tests yielded a SR increase of $>5$ percentage points, including two with a $\sim10$ percentage point increase, and the fourth time a SR increase of a few percentage points. We discuss possible reasons for these different performances in Sections $\S$\ref{sec:Elevation} and $\S$\ref{sec:OtherEffects}. The NIRC2 PSF images corresponding to before and after each of our closed-loop tests are shown in Figure~\ref{fig:CL_PSFs}. We can clearly see an improvement in the PSF shape, in particular the removal of a trefoil shape in the PSF from the first two closed-loop tests. A more detailed tracking of the PSF SR values throughout the closed-loop tests are shown in Figures~\ref{fig:ClosedLoop_SRs_02}~-~\ref{fig:ClosedLoop_SRs_13}.

\begin{table*}
    \centering
    \begin{tabular}{c|c|c|c|c|c|c}
        % &  &  &  &  &  & Total Corrections \\
        Date (UT) & Star & Pre Closed-Loop SR & Post Closed-Loop SR & $\Delta$SR & $\Delta$RMS & Total Corrections Applied \\
        \hline
        2023-08-02$^*$ & SAO68615 & 35.6\% $\pm$ 3.2\% & 41.1\% $\pm$ 2.6\% & 5.5\% $\pm$ 4.1\% & 130 nm &190 nm RMS \\ % & low-elevation \\
        2023-08-02$^{\dagger}$ & HD221914 & 49.8\% $\pm$ 3.0\% & 59.9\% $\pm$ 3.1\% & 10.2\% $\pm$ 4.3\% & 149 nm & 182 nm RMS \\ % & high-elevation \\
        2023-08-04 & SAO69743 & 57.4\% $\pm$ 2.9\% & 67.1\% $\pm$ 1.7\% & 9.7\% $\pm$ 3.4\% & 136 nm & 218 nm RMS \\ % & \\
        2023-08-13 & HIP83083 & 60.7\% $\pm$ 4.1\% & 64.0\% $\pm$ 4.8\% & 3.3\% $\pm$ 6.3\% & 79 nm & 128 nm RMS \\ % & different pupil rotation angle\\
    \end{tabular}
    \caption{Strehl Ratio differences and corresponding changes in RMS wavefront between NIRC2 images taken with the nominal primary mirror shape and the vZWFS-generated primary mirror shapes after the closed-loop. The last column shows the total RMS of the piston commands (in wavefront OPD) sent to the primary mirror segments by the end of each vZWFS closed-loop test. $^*$Telescope pointing at low-elevation. $^{\dagger}$Telescope pointing at high-elevation.}
    \label{tab:delta_SRs}
\end{table*}

\new{On a typical night, K-band SRs are 50-60\%. The three largest terms in the Keck AO error budget \citep{Wizinowich23} are: (1) bandwidth error, which is due to the time-lag in the AO loop between the measurement and the correction, (2) fitting error, where the number of actuators on the DM in the AO system limit the correction of atmospheric turbulence, and (3) the ``margin" error, which is a general term for unknown sources of aberrations added to match the measured SRs. \cite{Ragland22} demonstrated that this term is largely due to the primary mirror segment co-phasing errors. Taking into account the bandwidth and fitting errors, we are limited to SRs of ~75\% even if the ZWFS perfectly removed the ``margin" error by correcting the primary mirror segment phasing. Therefore, we expect the SRs to be limited to $<75\%$.}

%%%%%%%%%%%%%%%%%%%%%%%%%%%%%%%%%%%%%%%%%%%%%%%%%%%%%%%%%%%%%%%%%%%%%%
%%%%%%%%%%%%%%%%%%%%%%%%%%%%%%%%%%%%%%%%%%%%%%%%%%%%%%%%%%%%%%%%%%%%%%
\section{Analysis}

%%%%%%%%%%%%%%%%%%%%%%%%%%%%%%%%%%%%%%%%%%%%%%%%%%%%%%%%%%%%%%%%%%%%%%
\subsection{Segment Piston Measuring Stability}
\begin{figure*}
    \centering
    \includegraphics[width=0.8\textwidth]{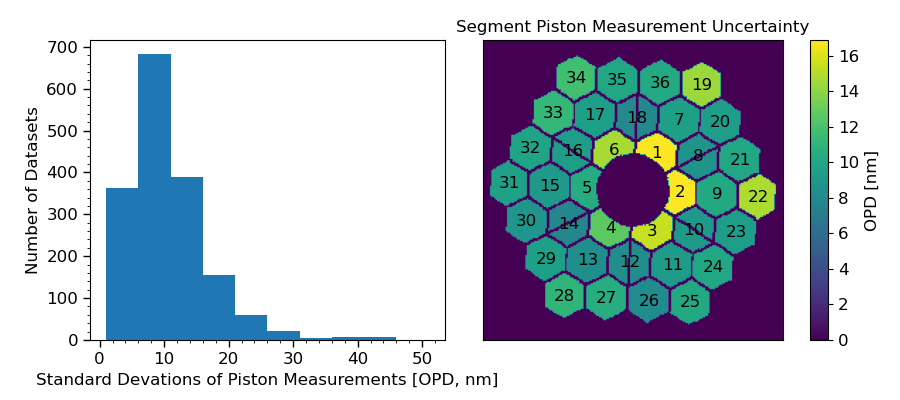}
    \caption{\textit{Left}: Histogram of all dataset standard deviations in segment piston measurements. The average uncertainty on a segment piston measurement is 11~nm. \textit{Right}: Map shows the average uncertainty for each segment.}
    \label{fig:meas_stdevs}
\end{figure*}

For each iteration of the closed loop sequences, we took five sets of vZWFS measurements (30-seconds integration time each), reconstructed the phases and extracted the segment piston values of each set, and then averaged each segment piston value over the five measurements. Figure \ref{fig:meas_stdevs} shows the distribution of standard deviations for all segment piston measurements in the datasets from all of the closed-loop tests. The average extracted piston value uncertainty is 11~nm (OPD). The map shows the average standard deviation for each segment, showing that we have larger uncertainties for the inner ring of segments, which are partially blocked by the central obscuration of the telescope top end.

%%%%%%%%%%%%%%%%%%%%%%%%%%%%%%%%%%%%%%%%%%%%%%%%%%%%%%%%%%%%%%%%%%%%%%
\subsection{Mirror Phasing at Different Elevations} \label{sec:Elevation}

\begin{figure*}
    \centering
    \includegraphics[width=0.8\textwidth]{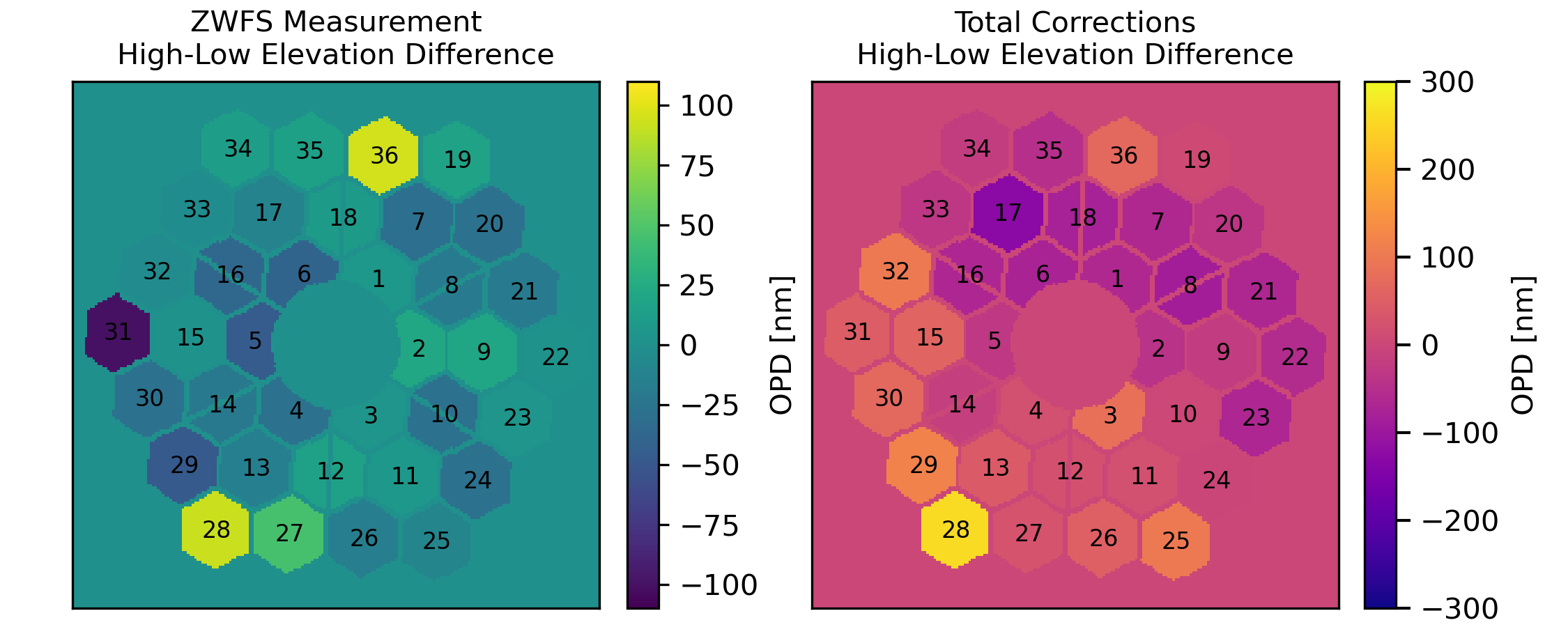}
    \caption{\textit{Left}: Difference between the vZWFS measurements of segment piston values with the telescope pointing at high and low elevations. These measurements were made before running the closed-loop, so with the primary mirror in its nominal shapes. \textit{Right:} Difference between the total piston commands sent to the primary mirror during the closed-loop runs at high and low elevations. We can see a global tip/tilt in the difference map of the total corrections applied.
    }
    \label{fig:HighLowDiffs}
\end{figure*}

We ran the closed-loop segment control at two different telescope elevations. It is known that segments enter a ``staircase" or ``terrace" mode when the telescope is pointing at low-elevations, which impacts PSF quality on NIRC2 \citep{Ragland18}. This effect is caused by spurious edge sensors readings as a result of the combination of sensor misalignments and segment rotation as gravity changes~\citep{Chanan23}. A lookup table of offsets for the desired edge sensor readings as a function of telescope elevation was generated and recently updated. We compare two closed-loop tests conducted on the same night (August 2nd, 2023) at different telescope elevations. The first closed-loop was conducted at a low-elevation of $\sim40$ degrees and the second closed-loop was conducted at a high-elevation of $\sim80$ degrees. The high-elevation closed-loop test yielded a SR increase on NIRC2 of 10.18 $\pm$ 4.25 percentage points, while the low-elevation before and after closed-loop comparison yields a 5.46 $\pm$ 4.13 SR percentage point increase. Figure \ref{fig:HighLowDiffs} shows the difference between the vZWFS measurements of the primary mirror in its nominal shape (prior to the vZWFS closed-loop iterations) at the low and high elevations. We closed the loop at both of these elevations, so we also show the difference in the total corrections applied to the primary between the closed-loop run at high and low elevations. In this map, we can more clearly see a global tip/tilt pattern difference.

%%%%%%%%%%%%%%%%%%%%%%%%%%%%%%%%%%%%%%%%%%%%%%%%%%%%%%%%%%%%%%%%%%%%%%
\subsection{High-Spatial Scale Structures}
\label{sec:HighSpatialScale}

\begin{figure}
    \centering
    \includegraphics[width=0.45\textwidth]{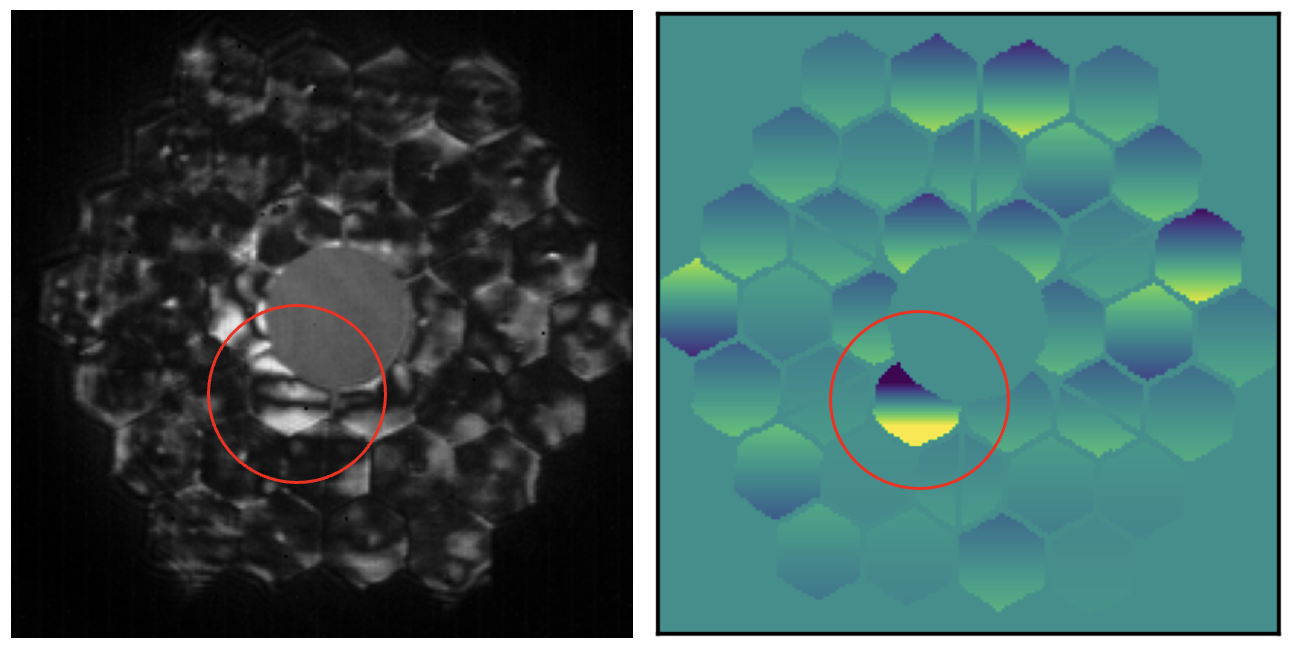}
    \caption{Segment Tilt seen in Zernike WFS image and in reconstructed segment phase.
    }
    \label{fig:TiltSeg}
\end{figure}

\begin{figure}
    \centering
    \includegraphics[width=0.45\textwidth]{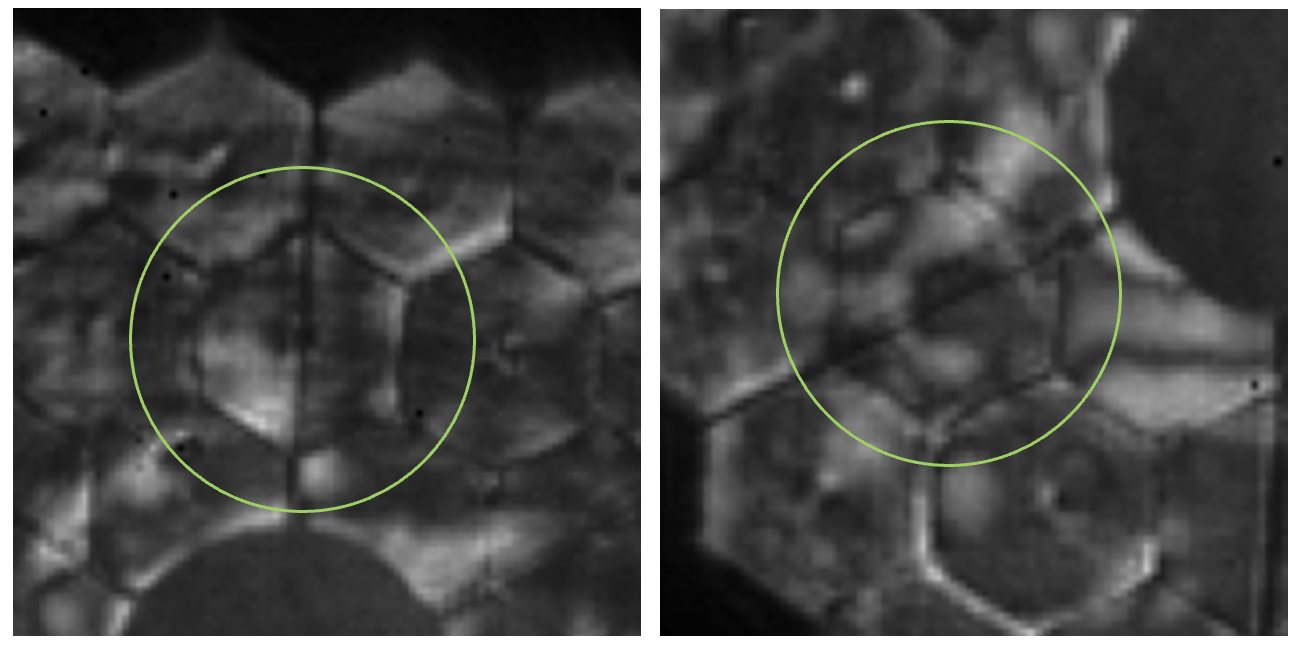}
    \caption{\textit{Left:} Low-wind effect seen on a segment with spider cutting across it. \textit{Right}: Higher order structure.
    }
    \label{fig:HighOrder}
\end{figure}

The spatial scale of the reconstructed wavefront is only limited by the number of pixels in the vZWFS pupil images. This study focused on segment piston, but future work will also test segment tip and tilt closed-loop control, as well as higher order aberration measurements such as segment warping. Fringing can be seen on certain segments, in particular Segment 4 in Figure \ref{fig:TiltSeg}, which is recovered in the phase reconstruction as segment tilt. A phase step can also be seen in some segments across which lies a spider, known as the low-wind effect, shown in Figure \ref{fig:HighOrder}. The dot in the center of each segment is where the central support post was located during segment fabrication.

Correcting segment tip/tilt, in addition to piston, is through the same ACS system controlling the three actuators behind each segment, and will be possible to implement in the same way. However, correcting any segment warping, is not possible to do at Keck in a remote, automated way; however it would be possible for the ELTs. One possible solution is to apply the correction on the DM, though with 2-3 actuators across each Keck segment, high-order structures cannot be corrected. After the High order Advanced Keck Adaptive optics (HAKA; \citealt{Wizinowich23}) upgrade, which consists of replacing the current \new{21$\times$21} DM by a high-order DM with $>6$ times more actuators, more high-spatial scale structures will be correctable with the Keck DM. The vZWFS could potentially be used to determine the required warping forces after segment exchanges.

%%%%%%%%%%%%%%%%%%%%%%%%%%%%%%%%%%%%%%%%%%%%%%%%%%%%%%%%%%%%%%%%%%%%%%
\subsection{Disentangling Primary Mirror Phasing from AO correction} \label{sec:OtherEffects}

Closed-loop control of the primary mirror segments by the vZWFS improves the image quality, as shown in this work for the first time with improvements in the NIRC2 PSF Strehl ratios. In this study, wavefront aberrations measured by the vZWFS were corrected using the primary mirror \new{segments}. However, this assumes the AO system does not introduce any aberrations or correct for any piston over segments and attributes all remaining aberrations to the primary mirror segments. In reality, there are AO residuals and other static aberrations. A follow-up study will be needed to carefully disentangle the sources of the aberrations corrected during our closed-loop operations. A more ideal operation, taking full advantage of the vZWFS's sensitivity, would include applying the correction at the source of the aberrations: correcting primary co-phasing errors with the primary mirror segments, and correcting AO residuals and other static aberrations with the Keck AO deformable mirror. 

Although the SHWFS does not sense phase-discontinuities, such as piston errors between the segments, if a SHWFS sub-aperture straddles two segments, it could sense the piston difference between the two segments as a local tip/tilt and send commands to the DM to correct it. Therefore, the alignment of the pupil with respect to the SHWFS sub-apertures and DM actuators will impact the vZWFS measurements of segment pistons. The follow-up study will consist of tests and analysis of the full AO telemetry in order to extract any primary mirror co-phasing errors potentially being corrected by the SHWFS AO system. Running our vZWFS closed-loop tests at different pupil rotation angles will be needed to disentangle this effect. In addition, running the vZWFS with the PyWFS closing the AO loop instead of the SHWFS will also provide a means for determining the effect of the SHWFS sub-apertures on the post-AO vZWFS measurement, while introducing the effect of PyWFS pixels.

%%%%%%%%%%%%%%%%%%%%%%%%%%%%%%%%%%%%%%%%%%%%%%%%%%%%%%%%%%%%%%%%%%%%%%
%%%%%%%%%%%%%%%%%%%%%%%%%%%%%%%%%%%%%%%%%%%%%%%%%%%%%%%%%%%%%%%%%%%%%%
\section{Conclusion}

We controlled the Keck primary mirror segment pistons in closed-loop using the vZWFS installed in the KPIC instrument behind the Keck II AO system. The vZWFS \new{doubles the dynamic range compared to the scalar ZWFS and} measurements were done on the AO-corrected wavefront. We ran four closed-loop tests on three nights in August 2023. In parallel, we used the NIRC2 science camera to image the PSF and monitor the impact of our closed-loop primary mirror segment piston correction on the Strehl ratio. All four runs resulted in an increase in the Strehl ratio, with two resulting in a $\sim$10 percentage point Strehl ratio increase. \new{Taking into account the largest error terms limiting the Keck AO performance, we expect to be limited to SRs of $<75\%$.} We also measured the impact of telescope elevation on post-AO segment-phasing, and observed a global tip/tilt indicating a low-order change in AO corrected wavefront with telescope elevation. The average uncertainty on our segment piston vZWFS measurements is 11~nm in OPD with larger uncertainties on the piston values of the inner ring of segments relative to the rest of the segments.

Our next steps include: (i) controlling segment tip and tilt, in addition to piston, (ii) determin\new{ing} telescope elevation and time dependence on the vZWFS corrections, and (iii) disentangl\new{ing} which wavefront errors measured by the vZWFS are due to segment co-phasing errors versus other sources of aberrations such as AO systematics.  In addition to measuring segment piston, tip, tilt values, the vZWFS can measure high spatial-scale structures, either from segment warping shapes or AO residuals and systematics. As \new{one of} the most photon-efficient and sensitive WFS, the vZWFS can be a powerful tool used as a second stage WFS in addition to being used for segment co-phasing and monitoring the shape of the primary mirror in parallel with science observations. 

Such advances in segment phasing and wavefront control, contemporaneous with science observations, will be crucial for high-contrast imaging on the ELTs in the 2030s and for the Habitable Worlds Observatory. It will also lead to better science performance on Keck in the short-term.

%%%%%%%%%%%%%%%%%%%%%%%%%%%%%%%%%%%%%%%%%%%%%%%%%%%%%%%%%%%%%%%%%%%%%%
\begin{acknowledgments}
The authors wish to recognize and acknowledge the very significant cultural role and reverence that the summit of Maunakea has always had within the indigenous Hawaiian community. We are most fortunate to have the opportunity to conduct observations from this mountain.

This work is funded by the Heising-Simons Foundation. Funding for KPIC has been provided by the California Institute of Technology, the Jet Propulsion Laboratory, the Heising-Simons Foundation (grants \#2015-129, \#2017-318, \#2019-1312, and \#2023-4598), the Simons Foundation (through the Caltech Center for Comparative Planetary Evolution), and the NSF under grant AST-1611623. Part of this research was carried out at the Jet Propulsion Laboratory, California Institute of Technology, under a contract with the National Aeronautics and Space Administration (80NM0018D0004).

M.S. acknowledges support from the Keck Visiting Scholars Program (KVSP) to conduct some of the analysis of the vZWFS closed-loop test results. M.vK. acknowledges support from the KVSP to interface with the Keck II primary mirror ACS system. D.E. was supported by a NASA Future Investigators in NASA Earth and Space Science and Technology (FINESST) fellowship under award \#80NSSC19K1423. D.E. also acknowledges support from the KVSP to install the KPIC Phase II upgrades required for KPIC VFN. J.X. is supported by another FINESST award under \#80NSSC23K1434 and also acknowledges support from the KVSP to commission KPIC Phase II. 

\end{acknowledgments}

\begin{figure}
    \centering
    \includegraphics[height=0.4\textheight]{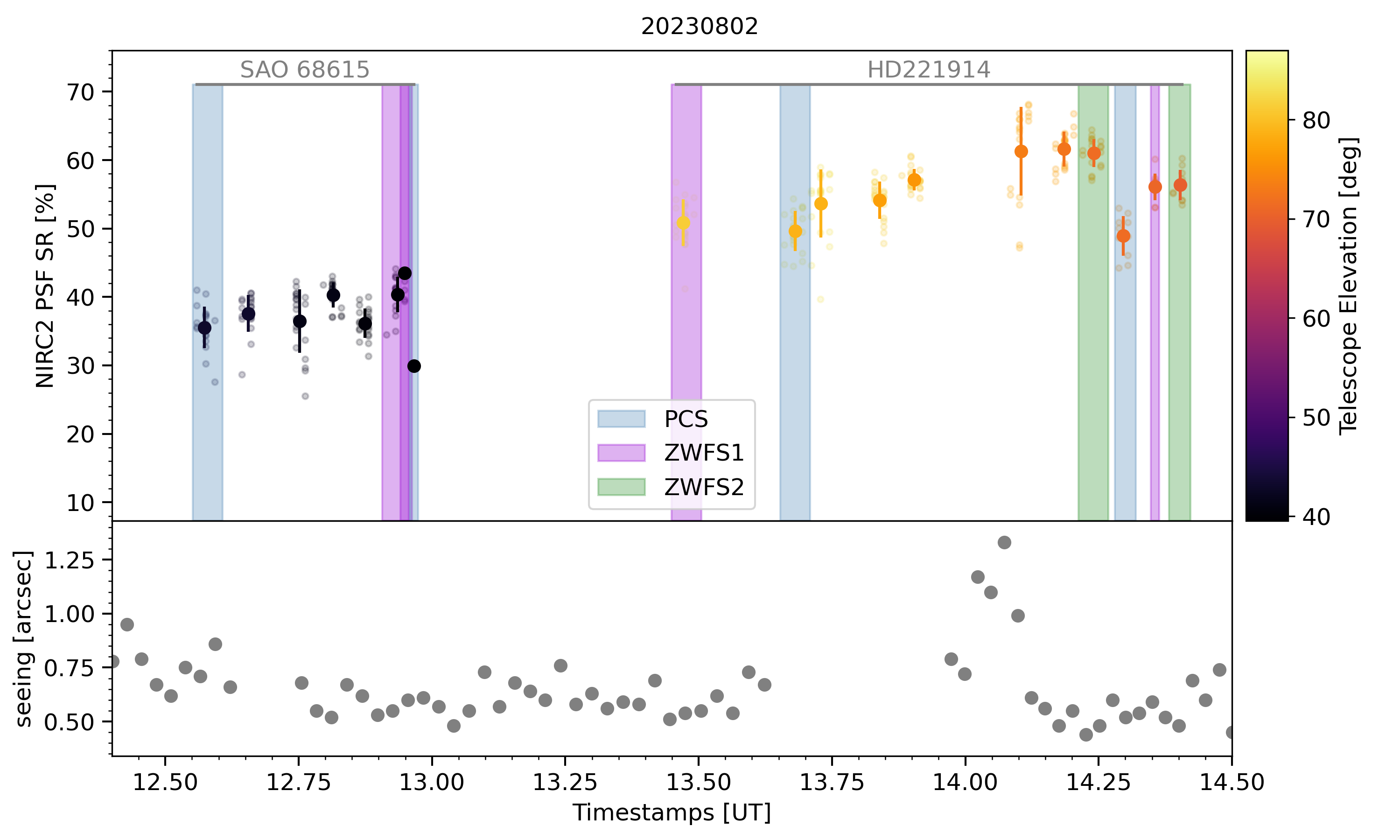}
    \caption{Summary of nightly measurements on August $2^{nd}$, 2023 (UT). The elapsed time during the first closed-loop test was 25~minutes and 35~minutes during the second closed-loop test (due to a pause during the spike in seeing). The blue shaded regions (PCS) correspond to the primary mirror being in its nominal shape. The purple (ZWFS1) and green (ZWFS2) shaded regions correspond to the segment piston offsets being applied as a result of the first (low-elevation) and second (high-elevation) vZWFS closed-loop tests, respectively.}
    \label{fig:ClosedLoop_SRs_02}
\end{figure}

\begin{figure}
    \centering
    \includegraphics[height=0.4\textheight]{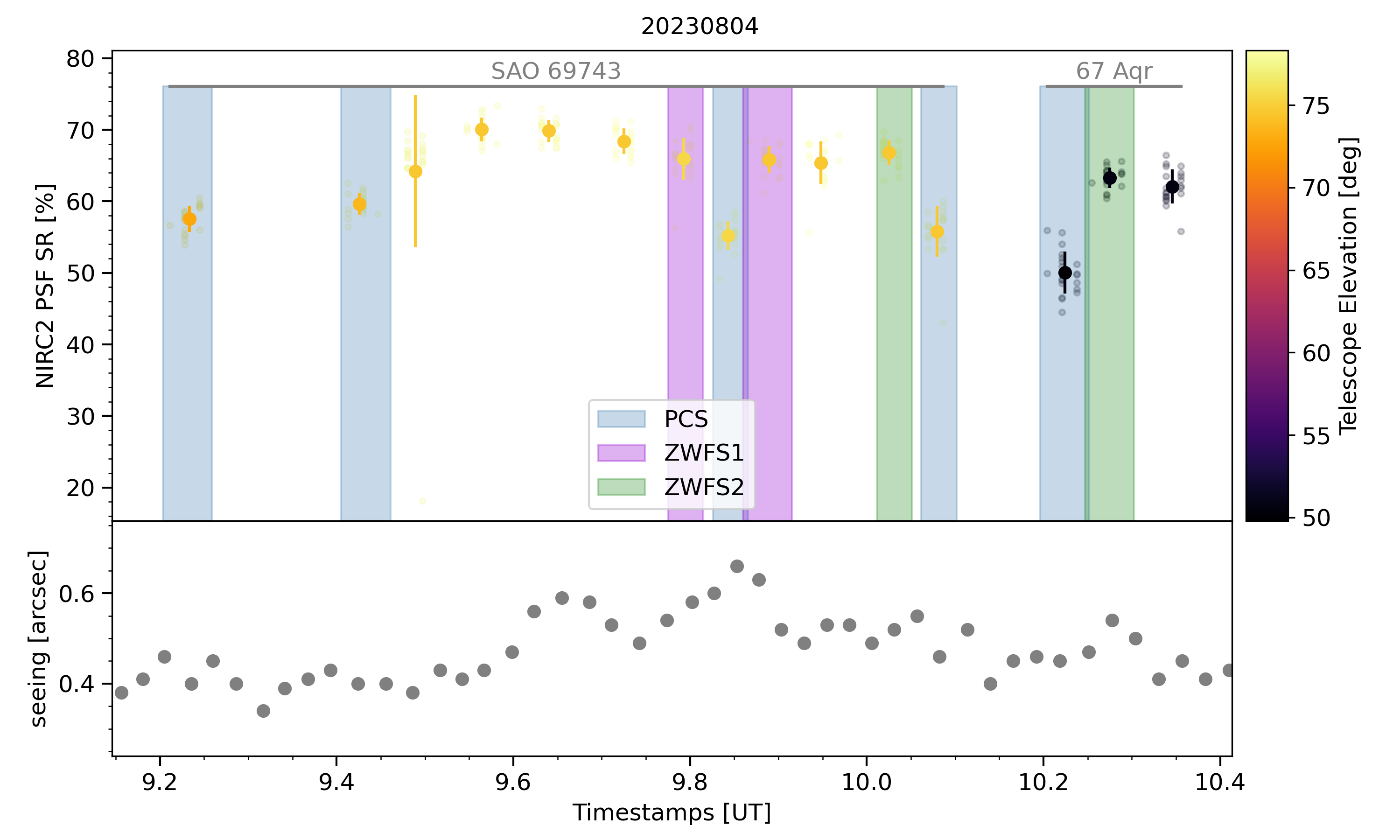}
    \caption{Summary of nightly measurements on August $4^{th}$, 2023 (UT). The elapsed time during the closed-loop test was 25~minutes. The blue shaded regions (PCS) correspond to the primary mirror being in its nominal shape. The purple (ZWFS1) shaded regions correspond to the segment piston offsets being applied as a result of the vZWFS closed-loop test. The green shaded region (ZWFS2) corresponds to two additional iterations of the closed loop test.}
    \label{fig:ClosedLoop_SRs_04}
\end{figure}

\begin{figure}
    \centering
    \includegraphics[height=0.4\textheight]{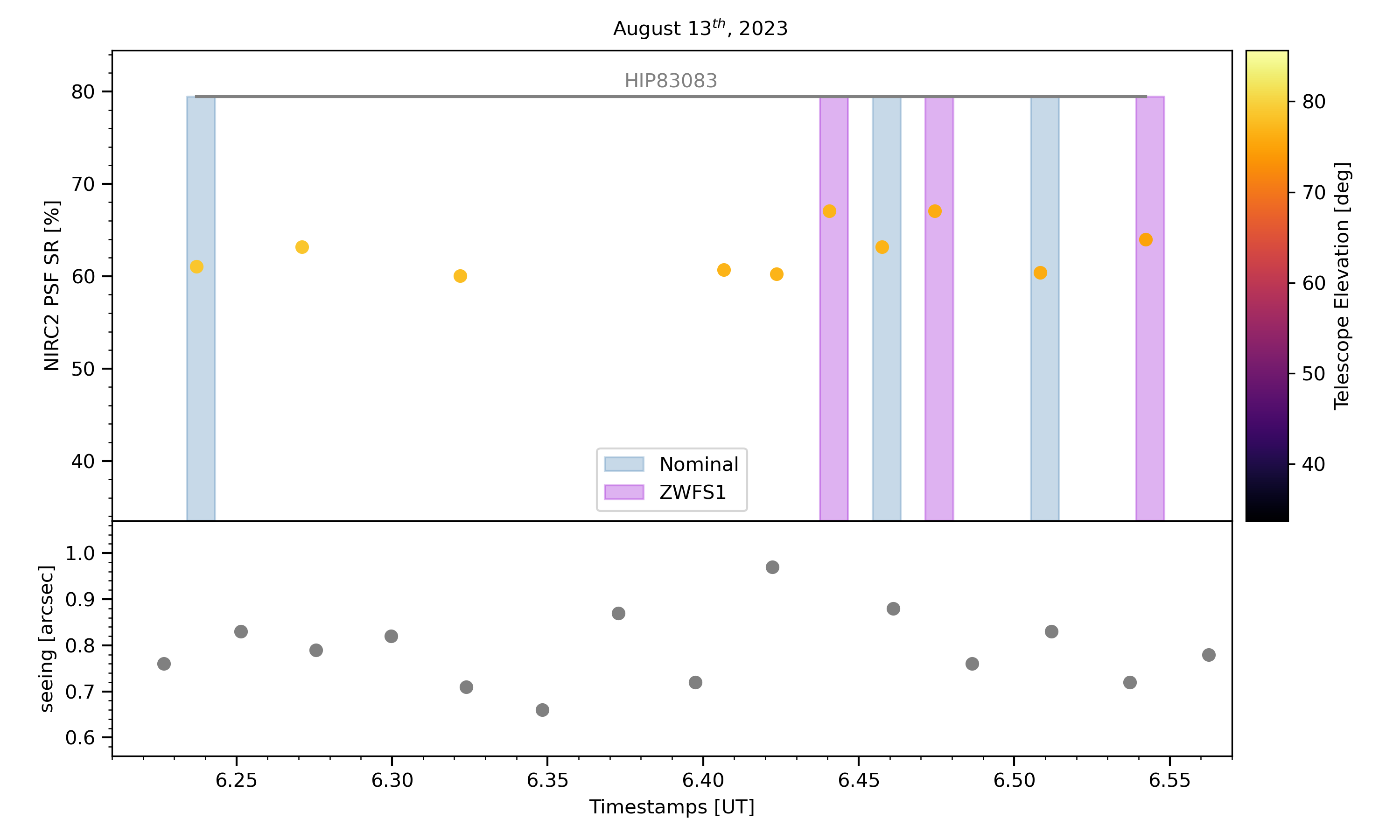}
    \caption{Summary of nightly measurements on August $13^{th}$, 2023 (UT). The elapsed time during the closed-loop test was 20~minutes. The blue shaded regions correspond to the primary mirror being in its nominal shape. The purple shaded regions correspond to the segment piston offsets being applied as a result of the vZWFS closed-loop test.}
    \label{fig:ClosedLoop_SRs_13}
\end{figure}

%%%%%%%%%%%%%%%%%%%%%%%%%%%%%%%%%%%%%%%%%%%%%%%%%%%%%%%%%%%%%%%%%%%%%%
\appendix
Nightly summary figures. Figures \ref{fig:ClosedLoop_SRs_02}, \ref{fig:ClosedLoop_SRs_04}, \ref{fig:ClosedLoop_SRs_13}.

%%%%%%%%%%%%%%%%%%%%%%%%%%%%%%%%%%%%%%%%%%%%%%%%%%%%%%%%%%%%%%%%%%%%%%
%%%%%%%%%%%%%%%%%%%%%%%%%%%%%%%%%%%%%%%%%%%%%%%%%%%%%%%%%%%%%%%%%%%%%%


\bibliography{sample631}{}
\bibliographystyle{aasjournal}



\end{document}